\documentclass[12pt,a4paper]{article}
\pdfoutput=1
\usepackage{jcappub}
\usepackage[english]{babel}
\usepackage{multirow}
\usepackage{ulem}
\usepackage{url}
\usepackage{caption}
\usepackage{subcaption}
\usepackage[utf8]{inputenc}

\def\bea{\begin{eqnarray}}
\def\eea{\end{eqnarray}}
\def\be{\begin{equation}}
\def\ee{\end{equation}}

\newcommand{\Eq}[1]{Eq.~\eqref{#1}}

\newcommand{\de}{\mathrm d}

\title{Upper limits on the dark matter content in globular clusters}

\author[a,b]{Javier Reynoso-Cordova,}
\emailAdd{javierisreal.reynosocordova@unito.it}
\author[a,b]{Marco Regis, and}
\author[b]{Marco Taoso}
\affiliation[a]{Dipartimento di Fisica, Universit\`{a} di Torino, via P. Giuria 1, I--10125 Torino, Italy}
\affiliation[b]{Istituto Nazionale di Fisica Nucleare, Sezione di Torino, via P. Giuria 1, I--10125 Torino, Italy}

\abstract{
We present a systematic analysis on the possible presence of dark mass components inside globular clusters (GCs). A spherical Jeans analysis is applied to the stellar kinematics of 10 nearby GCs. On top of the mass distribution provided by the luminous stellar component, we add either dark matter (DM), described by an NFW or Burkert mass profile, or an intermediate mass black-hole (IMBH), described by a point-like mass. Their existence would have important implications in the context of indirect DM searches. After profiling over the stellar parameters, we find no evidence neither for DM nor for IMBH. Upper limits on the two components are reported.
}

\date{\today}
\begin{document}
\maketitle

\section{Introduction}
\label{sec:Intro}
Globular clusters (GC) are old, very compact, gravitationally bound systems of stars.
They have a stellar mass of a few $\times\, 10^5\,M_\odot$, similar to the one of faint dwarf spheroidal (dSph) galaxies. On the other hand, GCs have a much higher density, with the stellar mass confined within a region of tidal radius of a few tens of parsec.
DSph galaxies are instead significantly more extended and they require the presence of a dominant dark matter (DM) component to explain their dynamics.
This is not the case for GCs, where there is no compelling evidence, at the time of writing, that they host a dark component different from dark remnants from stellar evolution.
However, the possible presence of a subdominant DM component is a very important question to address. It is indeed crucial to understand GC formation, and, in particular, whether GCs formed from a host halo which has been then (partially) stripped away, or if GCs are simply big star clusters.
Recent simulations of the formation and evolution of GCs in DM halos can be found in~\cite{Ardi:2020hlx,Wirth:2020dmb,Carlberg:2021qgo,Boldrini:2019saf,Ma:2019krp,Madau:2019srr} and references therein. The presence of a DM halo in GC could have dramatic consequences for indirect searches of particle DM. Indeed, several GCs in the Milky Way are very close to our location, and being very compact, they could have a very high $J$-factor, which would make them a prime target in the hunt for annihilating DM~\citep{Zaharijas:2007ci,2008ApJ...678..594W,2011ApJ...735...12A,Feng:2011ab,Bartels:2018qgr,Brown:2018pwq,2019arXiv190708564B,2021JCAP...02..010R}.

In this work, we analyze the line-of-sight (LOS) velocity of stars belonging to GCs, in order to infer the GC mass components.
In the past years, several works, involving different methods, have been devoted to the description of the dynamics of GCs and their mass profiles, see e.g., the seminal work in~\citep{1997AJ....114.1074M} and the discussion of the developments in \cite{2018MNRAS.478.1520B,H_nault_Brunet_2018}.  Recent measurements are now offering the possibility to analyse several thousands of stars for tens of GCs, greatly enhancing the statistical capability. Here we perform a systematic study involving ten GCs and using most updated data, taken from \cite{2018MNRAS.478.1520B} and \cite{2018MNRAS.473.5591K}.

We investigate the possible presence of a DM mass component, with the spatial density distributed following a NFW~\citep{1996ApJ...462..563N} or a Burkert~\cite{Burkert:1995yz} profile.
Additionally, we test the possibility that GCs host an intermediate mass black hole (IMBH) at their center. This would constitute a dark ``point-like'' source of mass, different from the extended distribution considered for DM.
As for the case of DM, the literature includes hints for a presence of IMBH in GCs~\citep{2008ApJ...676.1008N,2017MNRAS.464.2174B,2017Natur.542..203K}, as well as bounds~\citep{vanderMarel:2009mc,2019MNRAS.482.4713Z,2019MNRAS.488.5340B,Freire:2017mgu,Mann:2018xkm,Abbate:2019qoc,2018MNRAS.481..627A}, and this work offers a systematic up-to-date address to the topic.

The paper is organized as follows.
Section \ref{sec:data} describes the data on the LOS velocity of stars employed in our work, and presents the Jeans analysis adopted for the theoretical prediction.
In Section \ref{sec:Res}, we describe the statistical analysis and report the results in terms of bounds on the presence of DM and IMBHs in GCs.
We summarize consequences and conclusions in Section \ref{sec:Conc}.

\section{Data and methods}
\label{sec:data}

The analysis performed in this paper is based on two databases of LOS velocities: the sample collected by \cite{2018MNRAS.478.1520B} and recent measurements from the MUSE collaboration~\citep{2018MNRAS.473.5591K}. 
The sample from \cite{2018MNRAS.478.1520B} is a compilation of homogeneous measures from spectra obtained at the Very Large Telescope and Keck telescope, complemented by published measurements collected from the literature. The reported velocities typically cover three decades in radius and have an uncertainty around 0.5 km/s. Details on the sample selection and data analysis can be found in \cite{2018MNRAS.478.1520B} and \cite{2019MNRAS.482.5138B}, see also \cite{2021MNRAS.505.5957B} for a recent update.

The dataset collected by the MUSE integral-field spectrograph~\citep{2018MNRAS.473.5591K} greatly enlarges the number of stars acquired in each GC. Moreover, it provides LOS velocities at radii typically closer to the GC centres, allowing a more accurate derivation of the mass density in the innermost region.

\begin{table}[tb]
\begin{center}
\begin{tabular}{||c | c | c | c | c | c | c | c||} 
 \hline
 GC & RA & DEC  & D  & $R_{\rm{half}}$  & Mass & \# stars [B] &  \# stars [M] \\ [0.5ex] 
 \hline\hline
 47 Tuc & 6.02 & -72.08 & 4.52 & 4.10  & 8.95  &3709 & 12666\\ 
 \hline
 NGC 1851 & 78.53 & -40.05 & 11.95 & 1.95 & 3.18 & 656 & 5559\\
 \hline
 NGC 2808 & 138.01 & -64.86 & 10.06 & 2.17 & 8.64 & 1523 & 3431 \\
\hline
NGC 3201 & 154.4 & -46.41 & 4.74 & 3.33 & 1.60 & 2898 & 3283 \\
\hline
 $\omega$-cen & 201.70 & -47.48 & 5.43 & 7.93 & 36.40 & 2747 & 16282 \\
 \hline
 M 80 &  244.26 & -22.97 & 10.34 & 2.01 & 3.38 & 434 & 1151 \\
 \hline
 M 22 & 279.10 & -23.90 & 3.30 & 3.52 & 4.76 & 940 & 2465 \\
 \hline
 NGC 6752 & 287.71 & -59.98 & 4.13 & 2.51 & 2.76 & 1578 & 4580 \\
 
 \hline
 M 2 & 323.36 & -0.82 & 11.69 & 2.88 & 6.20 & 511 & 5238 \\
 \hline
 M 30 & 325.09 & -23.18 & 8.46 & 2.9 & 1.40 & 800 & 4864 \\
 [1ex] 
 \hline
\end{tabular}
\end{center}
\caption{GC parameters used in the analysis. I. Globular Cluster identifier. II. Right ascension $[\deg]$. III. Declination $[\deg]$. IV. Distance [kpc]. V. Half-light radius as derived from the fit through {\sc pyGravSphere} [pc]. VI. Total Stellar Mass taken from~\citep{2020IAUS..351..451H} $[\times 10^5 M_\odot]$. VII. Number of analyzed stars from the dataset in~\cite{2018MNRAS.478.1520B}.  VIII. Number of analyzed stars from the MUSE dataset~\citep{2018MNRAS.473.5591K}. }
\label{tab:GC_info}
\end{table}

Among all the GCs surveyed in \cite{2018MNRAS.478.1520B} and \cite{2018MNRAS.473.5591K}, we select the ones having more than $10^3$ stars in both catalogues, in order to deal with large statistics, and a reported mass-to-light ratio larger than 1.5, since we are interested in testing the presence of a ``dark'' component on top of the luminous one. 
This results into ten GCs. After this choice, we then select, in each GC, only stars having a large membership probability (in details, $P> 0.01$ in \cite{2018MNRAS.478.1520B} and $P>0.5$ in \cite{2018MNRAS.473.5591K}, following their definition of membership probability).
Binary stars can influence the velocity dispersion and therefore bias the determination of the mass of the GC, see e.g.~\cite{2020ApJ...896..152R,2021MNRAS.508.4385A}. In order to mitigate this effect, we exclude binaries from the sample of stars for which multi-epoch observations are available (we impose $P_{\rm SNGL}> 0.05$ in \cite{2018MNRAS.478.1520B} and $P_{\rm variable}<0.8$ in \cite{2018MNRAS.473.5591K}, where $P_{\rm SNGL}$ and $P_{\rm variable}$ are respectively the binary and single probabilities estimated in those references). We find that this cut leads to a non-negligible reduction of the velocity dispersions, and in general a better agreement between the two samples we are considering. Finally, after applying these selection requirements, for each GC we combine the two samples from \cite{2018MNRAS.478.1520B} and \cite{2018MNRAS.473.5591K} in an single dataset, taking care of the possible double counting of stars included in both surveys. 
The properties of all GCs composing our sample are reported in Table \ref{tab:GC_info}.

We constrain the stellar surface density of the GCs using the results of~\cite{2019MNRAS.485.4906D}, where the radial number density profiles of a large sample of GCs have been determined using data from Gaia DR2, from the Hubble Space Telescope, and ground-based surface brightness measurements.

The data of the ten GCs of Table \ref{tab:GC_info} are compared to a model built from a spherical Jeans analysis.
The assumptions we employ are that GCs can be described through i) a collisionless Boltzmann equation, ii) in a steady-state, iii) with a spherical phase-space distribution function, iv) without rotation. \\
Clearly, none of these assumptions is exactly realized in nature. On the other hand, there are indications that they can be good approximations for GCs.
GCs are very old stellar systems, with age much larger than relaxation time, i.e., with high number of gravitational encounters between stars. Deviations from thermodynamic equilibrium in GCs are limited and, in this regime, the Jeans equation provides a good approximate description~\citep{2008BT}, despite it is devised for collisionless systems, whilst GCs are collisional.
The high densities of GC imply that they are more resistant to external tidal forces than extended objects, like, e.g., dwarf spheroidal galaxies. However, there are indications of tidal effects in the GC outskirt~\citep{2020A&A...637L...2P}. As described below, we will consider relatively small radii where tidal forces are likely to be negligible.
Rotation has been measured in some GCs, e.g., $\omega$-Centauri~\citep{2018MNRAS.473.5591K,2019MNRAS.485.1460S}. Its contribution to the total kinetic energy of a GC is of the order of a few \%~~\citep{2019MNRAS.485.1460S}, and it is particularly negligible at the center,
which is the most relevant region to search for an IMBH or a highly concentrated DM profile.
On the other hand, since we are searching for subdominant components (i.e. DM and IMBH), the impact on their final bounds might be sizable, and neglecting rotation is a simplifying assumption of our method.

The above theoretical considerations are encoded in the spherical Jeans equation \citep{1980MNRAS.190..873B,10.1093/mnras/200.2.361}:
\be
    \frac{1}{\nu(r)}\frac{\partial}{\partial r}\left(\nu(r)\sigma_{r}^2\right) +
    \frac{2\beta(r)\sigma_{r}^2}{r} = -\frac{GM(<r)}{r^2},\\
    \label{eq:jeans}
\ee
where $\sigma_r$ is the radial velocity dispersion of stars, $M(<r)$ is the cumulative GC mass as a function of radius $r$, $\nu$ is the stellar radial density profile, and $\beta$ is the velocity anisotropy.
We model the various ingredients and solve \Eq{eq:jeans} by using the non-parametric Jeans code {\sc GravSphere} \citep{2017MNRAS.471.4541R} through the python implementation {\sc pyGravSphere} \citep{2020MNRAS.498..144G}.
For the stellar tracer density, we sum three Plummer spheres \citep{10.1093/mnras/200.2.361}:

\be
\nu(r) = \sum_{j=1}^{3} \frac{3 M_j}{4\pi a_j^3}\left(1+\frac{r^2}{a_j^2}\right)^{-5/2}
\label{eq:plummer}
\ee
where $M_j$ and $a_j$ are free parameters, providing the mass and scale length of each individual component.
Let us mention that we do not identify the Plummer spheres with specific stellar populations. Our approach is to adopt a phenomenological effective function that provides a good description of the measured total stellar density profile. In Appendix~\ref{sec:King} we repeat our analysis adopting a different stellar distribution, namely the King profile~\cite{1962AJ.....67..471K}.

The velocity anisotropy $\beta(r)$ describes the orbital structure of the stellar system and it is modeled as:
\be 
\beta(r) = \beta_0 + \left(\beta_\infty-\beta_0\right)\frac{1}{1 + \left(r_a/r\right)^\eta}
\label{eq:beta}
\ee
where the free parameters are the ``inner'' anisotropy $\beta_0$, the ``outer''  anisotropy $\beta_\infty$, the transition radius $r_a$, and the index $\eta$ giving the sharpness of the transition.

The total mass is $M(r)=M_*(r)+M_{\rm{dark}}(r)$. We compute the stellar mass $M_*$ assuming the same shape as the light profile given by \Eq{eq:plummer} but with a free parameter $M_{\star}$ fixing the overall normalization (which corresponds to allowing a free mass-to-light ratio).
In the case of DM, the mass is computed by integrating the density distribution. We consider two options:
an NFW density profile~\citep{1996ApJ...462..563N}:
\be
\rho_{\rm{NFW}}(r)=\frac{\rho_s}{\left(\frac{r}{r_s}\right)\left( 1 + \frac{r}{r_s} \right)^2}\;,
\label{eq:nfw}
\ee
and a Burkert profile~\cite{Burkert:1995yz}, describing a cored DM distribution:
\be
\rho_{\rm{Bur}}(r)=\frac{\rho_s}{\left(1+\frac{r}{r_s}\right)\left( 1 + \left(\frac{r}{r_s}\right)^2 \right)}\;.
\label{eq:Bur}
\ee
In \Eq{eq:nfw} and \Eq{eq:Bur} the two free parameters $\rho_s$ and $r_s$ give the
normalization and the scale radius of the profile.
In the case of IMBH, we simply assume a point source of mass $M_{BH}$.
From the radial velocity dispersion derived by solving \Eq{eq:jeans}, we compute the LOS velocity dispersion by means of 
\be
    \sigma_{\text{LOS}}^2(R) = \frac{2}{\Sigma_*(R)}\int_R^\infty \left(1-\beta\frac{R^2}{r^2}\right)
    \frac{\nu(r)\sigma_r^2(r) r}{\sqrt{r^2-R^2}}\de r,
    \label{eqn:LOS}
\ee
where $\Sigma_*(R)$ is the tracer surface mass density at the projected radius $R$, derived from \Eq{eq:plummer}:
\be
\label{eqn:Sigmastar}
\Sigma_*(R) = \sum_{j=1}^{3} \frac{M_j}{\pi a_j^2}\left(1+\frac{R^2}{a_j^2}\right)^{-2}.
\ee
In addition to $\sigma_{\text{LOS}}$, we also fit data concerning the surface density profile of stars $\Sigma_*$, as well as the fourth-order moments of the velocity distribution, called virial shape parameters (VSP) \citep{2017MNRAS.471.4541R}
\bea 
v_{s1} &=& \frac{2}{5} \int_0^{\infty} G\,M\, \nu(r)\,[5-2\beta(r)]\, \sigma_r^2\, r\, dr = \int_0^{\infty} \Sigma_*(R) \,\langle v_{\rm LOS}^4 \rangle\, R\, dR
\label{eqn:vs1} \\
v_{s2} & = & \frac{4}{35} \int_0^{\infty} G\,M\, \nu(r)\, [7-6\beta(r)] \,\sigma_r^2 \,r^3 \,dr
 = \int_0^{\infty} \Sigma_*(R) \,\langle v_{\rm LOS}^4 \rangle \,R^3\, dR\;,
\label{eqn:vs2}
\eea
where the first term gives the model prediction, while the second one is the estimator obtained from data.

The VSP can alleviate the degeneracy $\beta-\rho$ between the anisotropy of the stellar component and the dark density, and they can be measured in this context thanks to the large number of stars considered for each GC.
One could attempt to address the $\beta-\rho$ degeneracy also by considering tangential velocities from proper motion data. However, the analysis of currently available data can introduce more caveats than $\sigma_{\text{LOS}}$, in particular a larger impact related to the description of GC rotation (that we are not including in our model), see, e.g. \cite{2021arXiv210910998E}. For the sake of simplicity, we restrict our analysis to LOS velocities.

\section{Results}
\label{sec:Res}

\begin{figure}
\centering
    \begin{subfigure}{\textwidth}
    \centering
    \includegraphics[width=1.\textwidth]{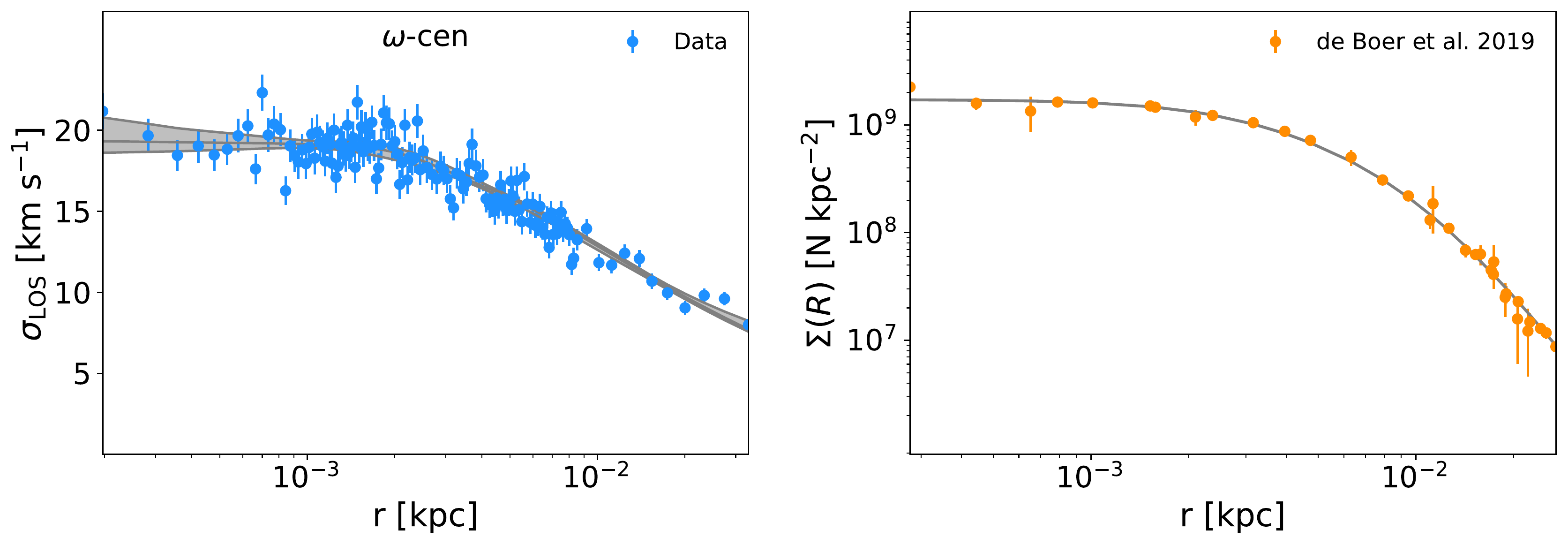}
    \end{subfigure} %
    \begin{subfigure}{1.\textwidth}
    \includegraphics[width=1.\textwidth]{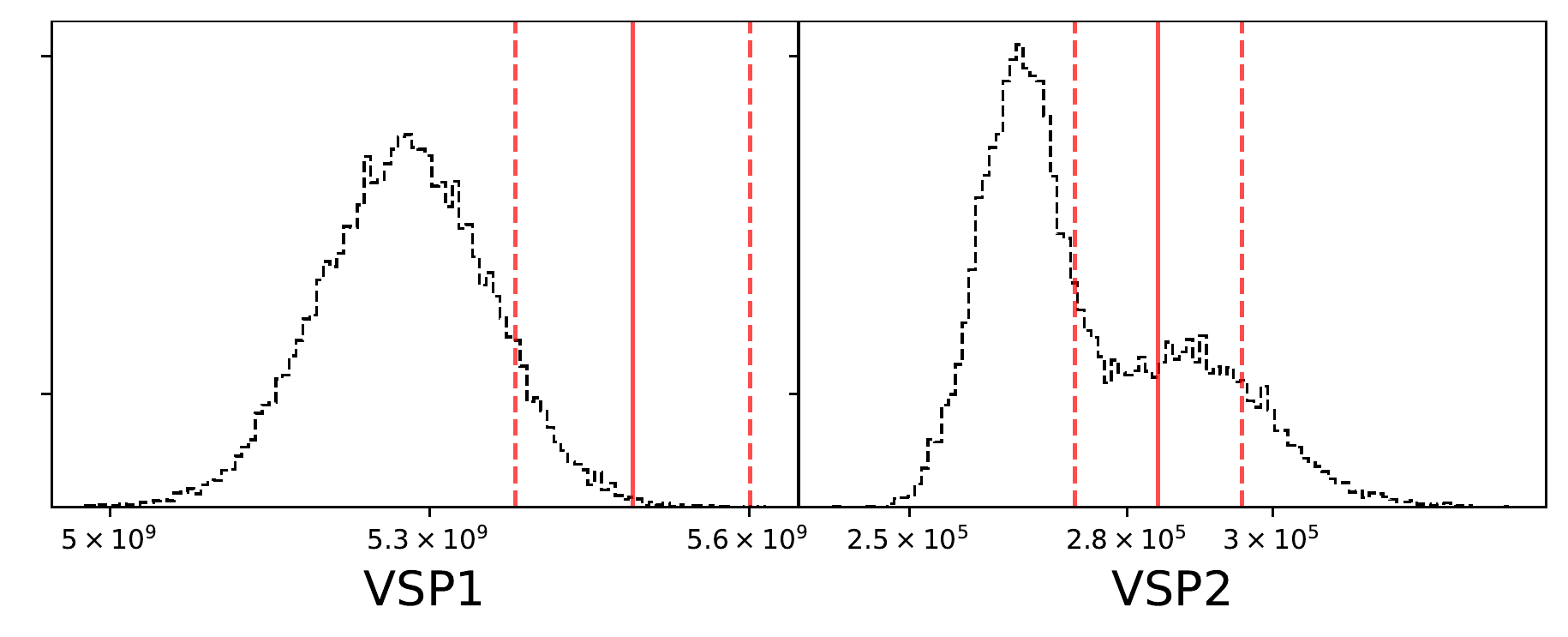}
   \end{subfigure}

   \caption{
   Analysis of the $\omega$-Centauri observables in the case of a mass model including an NFW DM profile.
   Left upper panel: line-of-sight velocity dispersion. Blue points correspond to the kinematic data obtained from the MUSE dataset~\citep{2018MNRAS.473.5591K} and from~\citep{2018MNRAS.478.1520B}. Right upper panel: the surface density profile derived from~\citep{2019MNRAS.485.4906D} (orange points). In both panels the gray region denotes the 95\% interval of the posterior distribution.        
      Lower panels: measurements of the VSP parameters (red solid line) and the corresponding 2-$\sigma$ intervals (dashed red lines) compared to the posterior distributions.}
   \label{fig:omega_cen_nfw}

\end{figure}

We compare the models with the observational data discussed in the previous section considering the likelihood function $\mathcal{L}$ given by:
\bea 
\label{eq:like}
-2\ln\mathcal{L}&=& \chi^2_{{\rm LOS}} + \chi^2_{\Sigma_*} + \chi^2_{{\rm VSP1}} +\chi^2_{{\rm VSP2}}=\chi^2.
\label{eq:likeM}
\eea
Our analysis includes measurements of the LOS velocity dispersion, the surface density profile and virial shape parameters.
The chi-squared $\chi^2$ for the LOS velocity dispersion is obtained using data binned over the projected radius $R$: 
\bea 
\label{eq:chi2}
\chi^2_{{\rm LOS}}  =  \sum_{j} \frac{ \left( \sigma_{{\rm LOS},j} - \bar{\sigma}_{{\rm LOS},j}\right)^2}{ \sigma_{\sigma_{{\rm LOS},j}}^2},
\eea
where the quantity $\bar{\sigma}_{{\rm LOS},j}$ is the observed velocity dispersion in the radial bin $j$. Following~\citep{2017MNRAS.471.4541R,2020MNRAS.498..144G}, we use a number of bins equal to $\sqrt{N_*},$ where $N_*$ is the number of stellar tracers. The error in each bin, $\sigma_{\sigma_{{\rm LOS},j}}$ includes Poissonian and sampling uncertainties, see~\citep{2017MNRAS.471.4541R,2020MNRAS.498..144G} for details. 
The dispersion velocity predicted by the model, $\sigma_{{\rm LOS},j}$, is computed as explained in the Section~\ref{sec:data}, see \Eq{eqn:LOS}.
Concerning the virial shape parameters, the chi-squared $\chi^2_{{\rm VSP1}}$ and $\chi^2_{{\rm VSP2}}$, are obtained analogously to \Eq{eq:chi2}, using the predicted values from the models computed with \Eq{eqn:vs1} and \Eq{eqn:vs2}, and the corresponding observational quantities and estimated errors, obtained as detailed in \citep{2017MNRAS.471.4541R,2020MNRAS.498..144G}:
\bea 
\label{eq:chi2vsp}
\chi^2_{{\rm VSP1}}  =  \frac{ \left( v_{s1} -\bar{v}_{s1}\right)^2}{ \sigma_{v_{s1}}^2} \;\;\;,\;\;\;\chi^2_{{\rm VSP2}}= \frac{ \left( v_{s2} -\bar{v}_{s2}\right)^2}{ \sigma_{v_{s2}}^2}.
\eea
Finally, we fit the surface density profiles of~\cite{2019MNRAS.485.4906D}:
\bea 
\label{eq:chi2Sigma}
\chi^2_{\Sigma_*}=\sum_{j} \frac{ \left( \Sigma_{*,j} - \bar{\Sigma}_{*,j}\right)^2}{ \sigma_{\Sigma_*,j}^2},
\eea
where $\bar{\Sigma}_{*,j}$ and $\sigma_{\Sigma_*,j}$ are respectively the surface number density in the radial bin $j$ and the corresponding error from~\cite{2019MNRAS.485.4906D}, while the prediction of the model $\Sigma_{*,j}$ is computed with \Eq{eqn:Sigmastar}.
In \Eq{eq:chi2Sigma}, for each GC, we include only measurements corresponding to the radial distances probed by the LOS velocity dispersion in \Eq{eq:chi2}. The reason is as follows.
At the largest radii considered in~\cite{2019MNRAS.485.4906D}, the background contamination from field stars becomes dominant.
This introduces a systematic effect from a region which is not particularly relevant for our purposes.
On the other hand, in the innermost radii and for some GCs, there is some level of mismatch between our model and the data. The same happens for the models considered in~\cite{2019MNRAS.485.4906D}. This might be due to the incompleteness of the sample of stars in these regions, or to the fact that the relative simple models adopted are not tailored to fit simultaneously both the outskirts and centres of the GCs. Again, to avoid the fit to be significantly affected by data from outside the regions were we perform the dynamical analysis, we exclude these internal regions from the surface density fit.

\begin{figure}
\centering
    \begin{subfigure}{1.\linewidth}
    \centering
    \includegraphics[width=1.\textwidth]{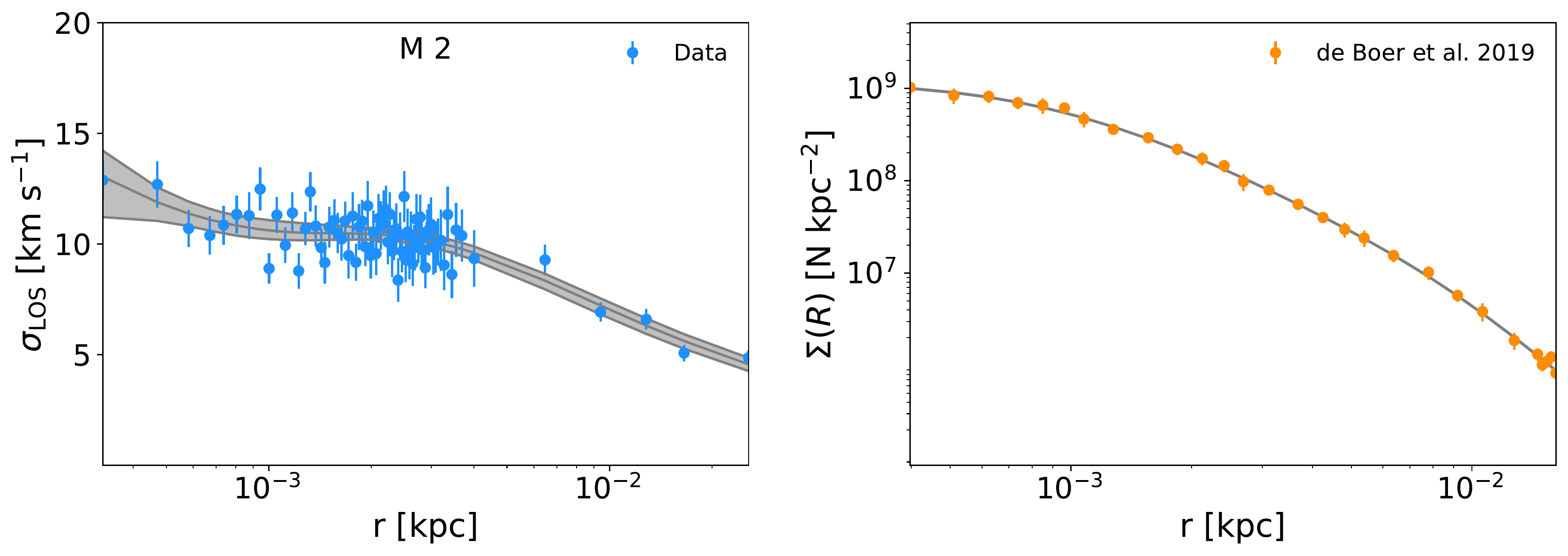}
    \end{subfigure} %


    \begin{subfigure}{1.\linewidth}
    \includegraphics[width=1.\textwidth]{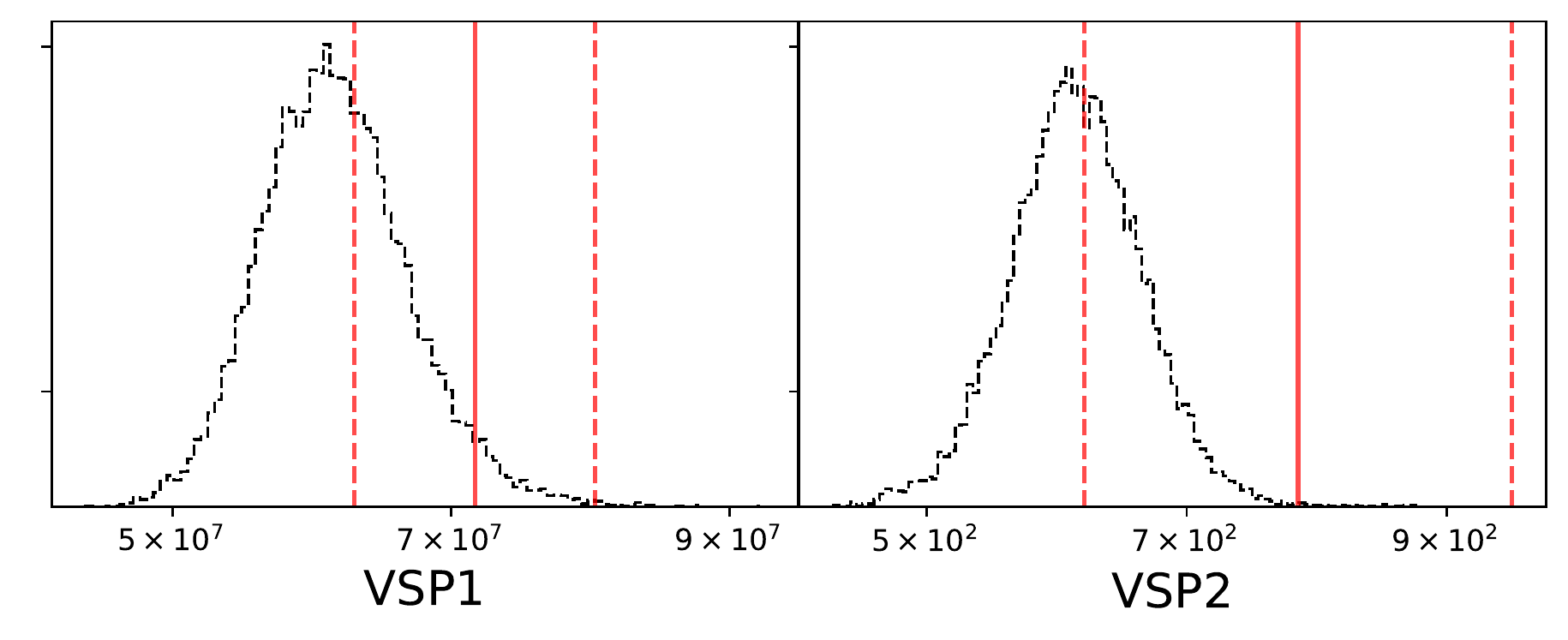}
   \end{subfigure}
   \caption{Same as Figure \ref{fig:omega_cen_nfw}, but for the M2 GC.
   }
    \label{fig:M2_nfw}

\end{figure}

The models outlined in Section~\ref{sec:data} contain 11 parameters describing the stellar system plus two (one) parameters for the DM (IMBH) component. We explore this parameter space with a Markov Chain Monte Carlo (MCMC) analysis employing the ensemble sampler Emcee~\cite{Foreman-Mackey:2012any}. We test different algorithms in order to ensure a good coverage of the parameter space and to verify the robustness of the results. For the plots presented here, we use 500 walkers and $10^4$ steps with 5000 burn in.

To sample the parameters, we use flat priors within the range reported in Table \ref{tab:priors}. 
We have 4 free parameters for the velocity anisotropy ($\beta_{\infty}$, $\beta_0$, $\eta$, $r_a$), a total of 7 parameters for the stellar profile ($m_1$, $a_1$, $m_2$, $a_2$, $m_3$, $a_3$, $M_\star$) and two extra parameters ($\rho_s$, $r_s$) in the DM case, while one extra parameter ($\rho_s$) in the IMBH case. It is worth noticing that, in the case of the stellar parameters, they are allowed to vary within a range around a previously fitted value. Namely, prior the MCMC, {\sc pyGravSphere} fits the surface density through a minimization process determining the associated best-fit values of the parameters of the stellar profile, and then, during the MCMC, they are allowed to vary, around the surface density best-fit, within the reported range. For the overall stellar mass $M_{\star}$ we consider the value reported in Table \ref{tab:GC_info} as the reference value, again allowing some variation during the MCMC analysis. Also, since the anisotropy parameter $\beta$ can take values from $-\infty$ to 1, {\sc GravSphere} uses the symmetrized anisotropy parameter \citep{2017MNRAS.471.4541R} $\hat{\beta}=\beta/(2-\beta)$
which takes a value of 1 for a full radial anisotropy and $-1$ for tangential anisotropy.
Note that, for the sake of generality and to check the consistency between the DM and BH findings, we allowed $r_s$ to take very small values, 
even though such extremely concentrated DM halos are not expected. Clearly, we cannot efficiently constrain profiles with $r_s$ much smaller than the radius of the first data-point (or said in a different way, DM profiles with extremely high concentration are allowed by data). The bounds on the DM mass are in any case independent from choice of the priors, as we will discuss later on.

\begin{table}[tb]
\begin{center}
\begin{tabular}{|| c | c | c ||} 
 \hline
 Parameters & min. value & max. value \\
 \hline\hline
 $\log_{10}(\rho_s/(M_\odot/\rm{kpc}^3))$ & 0 & 30 \\ 
 \hline
 $\log_{10}(r_s/\rm{kpc})$ & -8 & -1 \\
 \hline
 $\beta_\infty$ & -1 & 1 \\
 \hline
 $\beta_0$ & -1 & 1  \\
 \hline
 $\eta$ & 1. & 10.   \\
 \hline
 $r_a/\rm{kpc}$ &  $R_{\rm{half}}/10$ & $10\cdot R_{\rm{half}}$ \\
 \hline
 $M_j/M_\odot$ & $0.1(M_{j,\rm{bf}})$ & $1.9(M_{j,\rm{bf}})$  \\
 \hline
 $a_j/\rm{kpc}$ & $0.1(a_{j,\rm{bf}})$ & $1.9(a_{j,\rm{bf}})$   \\
 \hline
 $M_\star/M_\odot$ & $M_{\star}/2.$ & $2.5M_{\star}$   \\

 \hline
\end{tabular}
\end{center}
\caption{Priors used in the MCMC. The parameters of the stellar profile can vary around their best-fit values obtained by fitting the stellar surface density profile only, before running the MCMC. The total stellar mass $M_{\star}$ is allowed to vary around the value provided in Table \ref{tab:GC_info}.}
\label{tab:priors}
\end{table}

We choose two illustrative examples to show the typical outcomes of the analysis, focusing on the case with the DM component included in the fit.
In Fig.~\ref{fig:omega_cen_nfw} we show the case of $\omega$-Centauri, which is among the brightest and well studied GCs in the literature. We report the LOS velocity dispersion profiles, the surface density profiles, and the estimates of the virial shape parameters from the data, as well as the corresponding model predictions from the fit. As an additional example we show the same plots but for the GC M2 in Fig.~\ref{fig:M2_nfw}. The LOS velocity dispersion profiles for all the others GCs are in the Appendix~\ref{sec:AppendixA}.
In the Appendix~\ref{sec:AppendixB}, we show triangle plots reporting the posterior distributions of all the model parameters, including the stellar ones, for one example ($\omega$-Centauri).

The main goal of this work is to derive upper limits on the presence of a DM halo or an IMBH lying the center of GCs. In a Bayesian context, credible intervals are obtained from the posterior distribution. However, given that we find the parameters of the dark components, namely $\rho_s$ and $M_{\rm{BH}}$, to be unconstrained from below, care must be taken to deal with volume effects in the marginalization and to ensure that the results are independent on the choice of the priors. 
For the sake of robustness, we decide to adopt a frequentest approach, and derive our bounds performing a profile likelihood analysis. 

More specifically, we compute the likelihood ratio $\lambda_b=\mathcal{L}(\vec \theta_{\rm b.f.},\vec \Pi_{\rm b.f.})/\mathcal{L}(\vec \theta,\vec \Pi^\prime_{\rm b.f.})$, where $\vec \theta$ are the parameters of interest ($\rho_s$ and $r_s$ in the DM case, and $M_{\rm{BH}}$ in the IMBH case), $\vec \Pi$ are the stellar (nuisance) parameters, the suffix b.f. indicates the best-fit value, and the $^\prime$ symbol highlights that in the denominator we compute the best-fit values of $\vec \Pi$ at any given value of $\vec \theta$, so they are in general different from the global best-fit appearing in the numerator. It is easy to see that $-2\,\ln \lambda_b = \Delta \chi^2= \chi^2(\vec \theta,\vec \Pi^\prime_{\rm b.f.})-\chi^2(\vec \theta_{\rm b.f.},\vec \Pi_{\rm b.f.})$. The one-sided 95 \% C.L. exclusion limits on the two parameters $\rho_s$ and $r_s$ (or on the single parameter $M_{\rm{BH}}$) are obtained for $\Delta \chi^2>4.61 $ ($\Delta \chi^2>2.71).$ 

\begin{figure}[tb]
\centering
    \begin{subfigure}{1.\linewidth}
    \centering
    \includegraphics[width=1.\textwidth]{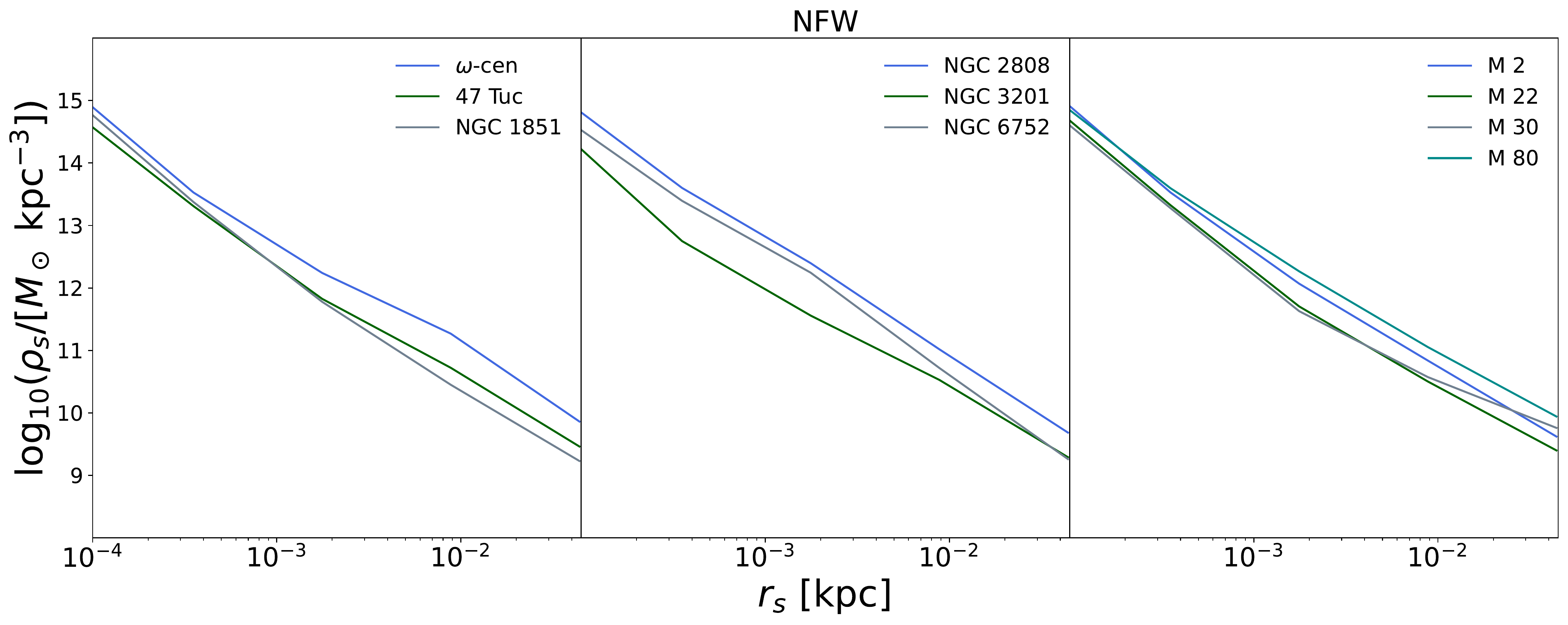}
    \end{subfigure} %


    \begin{subfigure}{1.\linewidth}
    \includegraphics[width=1.\textwidth]{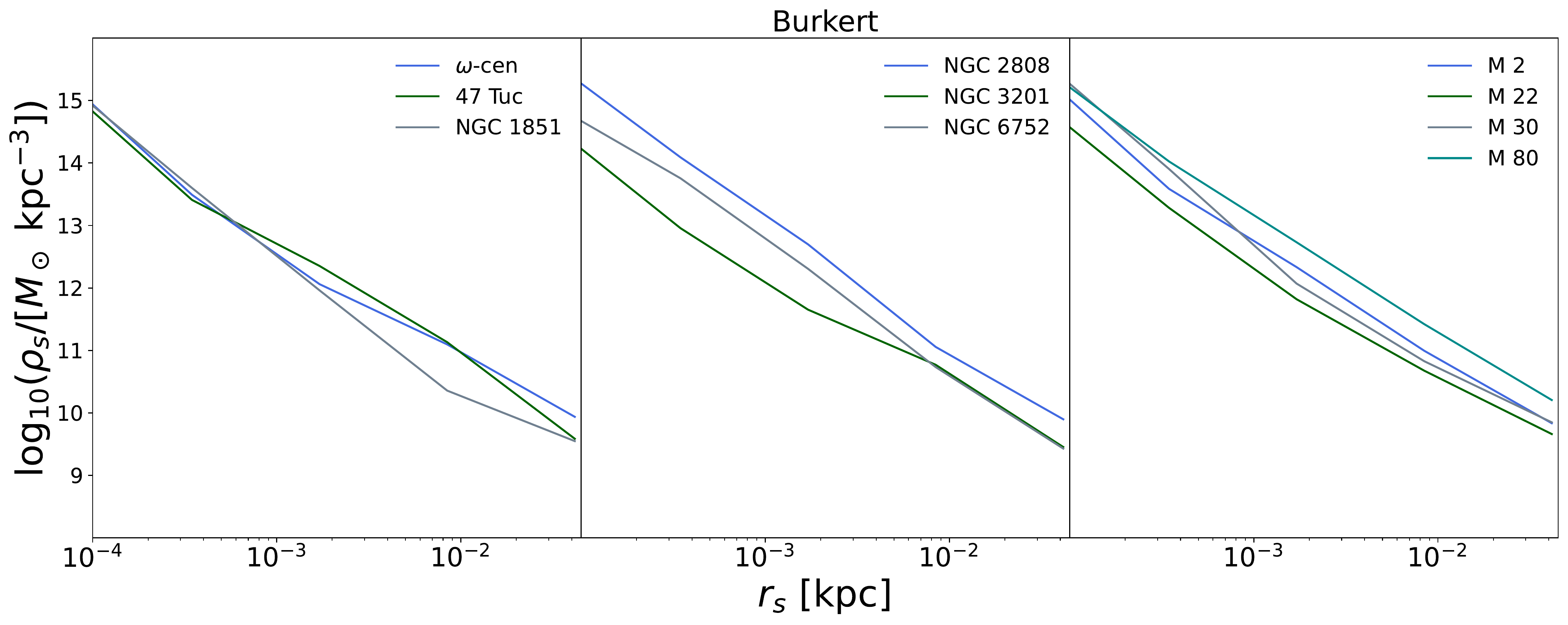}
   \end{subfigure}
\caption{Upper limits at 95\% C.L. on $\rho_s$ as a function of $r_s$ for an NFW (top panel) and Burkert (bottom panel) DM profile. The bounds are obtained following the profile
likelihood procedure described in Section \ref{sec:Res}, and the different panels show all the analyzed GCs.}
\label{fig:upperlimitsDM2} 
\end{figure}

\begin{figure}[tb]
\begin{center}
\includegraphics[width=1.\textwidth]{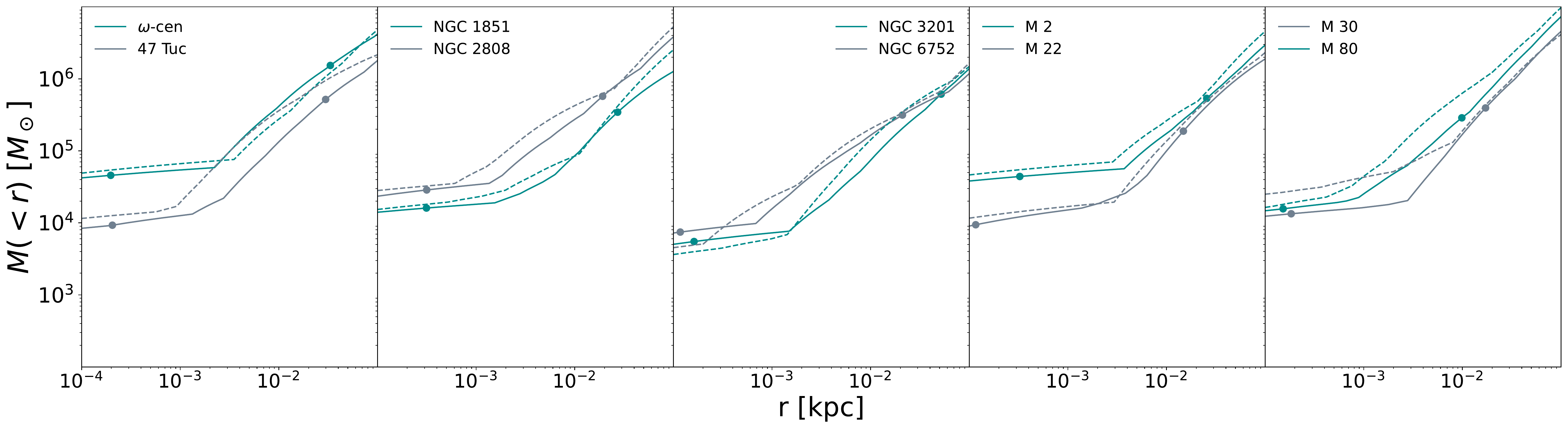}
\caption{Upper limits at 95\% C.L. on the DM mass enclosed within a radius $r$ for the different GCs analyzed. Solid lines are the C.L. obtained using and NFW profile, dashed for Burkert DM profile.
Dots correspond to the minimum and maximum radii covered by the measurements of the velocity dispersion.}
\label{fig:upperlimitsDM} 
\end{center}
\end{figure}

Results are shown in Figs.~\ref{fig:upperlimitsDM2}, ~\ref{fig:upperlimitsDM}, in Table~\ref{tab:chi2} (DM), and in Table~\ref{tab:BHbounds} (IMBH), and discussed below.
We note also that, for all the GCs analyzed, we do not find any statistical preference for DM or IMBH, namely we find $\mathcal{L}(\vec \theta_{\rm b.f.},\vec \Pi_{\rm b.f.}) \simeq \mathcal{L}(\vec \theta=0,\vec \Pi^\prime_{\rm b.f.})$.
We verified that the same conclusion can be drawn by looking at the marginalized posteriors of $\rho_s$ and $M_{\rm{BH}}$, which are compatible with zero.
In Table~\ref{tab:chi2}, we display the number of data-points entering the statistical analysis and report the  best-fit values of the $\chi^2$ obtained from Eq.~\ref{eq:likeM}, for models including an NFW DM halo. Similar values are obtained for the cases of a Burkert profile or an IMBH.

\subsection{Dark Matter}
\label{sec:darkmatter}

\begin{table}[tb]
\begin{center}
\begin{tabular}{|| c | c | c | c ||} 
 \hline
 GC & $N_{\rm KIN}$ & $N_{\rm S.D.}$ & $\chi^2_{b.f.}$ \\
 \hline\hline
 47 Tuc & 127 & 115 & 262    \\ 
 \hline
 NGC 1851 & 79 & 72& 131  \\
 \hline
 NGC 2808 & 70 & 60 & 141  \\
 \hline
 $\omega$-cen & 138 &33& 215  \\
 \hline
 M 80 &  40 & 134 & 120  \\
 \hline
 M 22 & 56 & 59 & 85  \\
 \hline
 NGC 6752 & 78 & 83 & 168  \\
 \hline
 M 2 & 76 & 28 & 72  \\
 \hline
 M 30 & 73 & 76 & 124  \\
\hline
 NGC 3201 & 71 & 51   & 126  \\
 \hline
\end{tabular}
\end{center}
\caption{II. Number of kinematic data-points included in our statistical analysis. III. Number of data points for the surface density. IV. $\chi^2$ corresponding to the best-fit for models including an NFW DM halo (similar values are obtained for the Burkert profile and the IMBH case).}
\label{tab:chi2}
\end{table}

In the top panel of Fig.~\ref{fig:upperlimitsDM2}, we show the 95\% C.L. upper limit on $\rho_s$ as a function of $r_s$ obtained from the profile likelihood procedure described above and assuming an NFW DM profile. Clearly, as the radius increases, the bound on the normalization of the profile decreases, since it is the mass distribution that is constrained by the kinematic analysis.

Fig.~\ref{fig:upperlimitsDM} reports the 95\% C.L. upper bound on the DM mass enclosed within a given GC radius $r$. It contains the same information as in Fig.~\ref{fig:upperlimitsDM2}, but shown in a different way.
We note that the increase in the mass bound is slower at low radii and steeper at large radii.
This can be understood as follows: for radius $\lesssim$ first kinematic point (which is the dispersion velocity at smallest observed radius), the profiles that can maximize the enclosed mass are the ones with $r_s\lesssim$ first kinematic point. For an NFW profile, the mass increases logarithmically for $r> r_s$ and so does the mass bound. 
It is easy to see that for sufficiently small $r_s$ (again $\lesssim$ first kinematic point) the bound is independent on $r_s$, so extending the prior range to very low values has no impact on the constrained mass.
When we consider larger radii, DM profiles that can maximize the allowed mass have larger $r_s$, and for $r_s\gtrsim r$, the mass of an NFW profile grows more steeply than $\ln r$, up to $\propto r^2$ for $r_s \gg r$. Clearly, the exact scaling also depends on the shape of the velocity dispersion profile and on the stellar mass profile, that can be somewhat different in different GCs.
In Fig.~\ref{fig:upperlimitsDM}, the minimum and maximum radius for the $\sigma_{\rm{LOS}}$ data points of each GC are reported.

Qualitatively, similar considerations hold for a Burkert DM profile. The upper limits on $\rho_s$ as a function of $r_s$ are shown in the bottom panel of Fig.~\ref{fig:upperlimitsDM2}. The bounds on the DM mass enclosed within a given GC radius, presented with dashed lines in Fig.~\ref{fig:upperlimitsDM}, are similar to the ones obtained for an NFW distribution.

\smallskip

In Fig.~\ref{fig:mass_to_light} in the Appendix~\ref{sec:AppendixA}, we show the posterior distributions of the Mass-to-Light ratios derived from the analysis with an NFW profile and for the different GCs. Very similar results are obtained considering a Burkert profile or an IMBH, since these dark components provide only a small contribution to the total mass.
In the same figure, we also report the values of the Mass-to-Light ratios presented in~\citep{hilker_baumgardt_sollima_bellini_2019} (red lines). In general, our estimates are in good agreement with those of~\citep{hilker_baumgardt_sollima_bellini_2019}, although a certain level of tension is present in few cases. Similar mismatches (at the level of a few tens of \%) are found in the literature, see e.g.~\cite{2015ApJ...812..149W,2005ApJS..161..304M,2009MNRAS.400..917L} and references therein, and are likely due to systematic uncertainties related to the modeling and measurements.

\bigskip
Recently, the possible presence of a DM halo in $\omega$-Centauri has been investigated in~\cite{2021arXiv210910998E,2019arXiv190708564B} on the basis of a Jeans analysis similar to the one performed here. That work analyzed the LOS dispersion profiles from~\cite{2018MNRAS.478.1520B} as well as measurements of the proper motion of stars. The upper bounds on the DM content for an NFW distribution are comparable to the one in Fig.~\ref{fig:upperlimitsDM}. However, Ref.~\cite{2021arXiv210910998E} finds significant evidence for the DM component. This different conclusion is probably mainly arising from a different treatment of the stellar component. In particular, our modeling include more free parameters to describe the stellar profile and velocity anisotropy.
Ref.~\cite{Carlberg:2021mpl} determined the velocity dispersion profiles of several GCs using Gaia EDR3 data. For NGC 6752 (and NGC 6205) it is reported an hint of a rising velocity dispersion at radii larger than $r\gtrsim30-50$ pc, which correspond to distances beyond those probed in our analysis. In principle, these results could be consistent with the presence of a DM halo. However, further investigation is needed to corroborate these findings, for instance to confirm the membership of the stars to the cluster. Previous analyses in the literature have not find evidence for DM in 47 Tucanae~\cite{Lane:2009fz}, in M22 and M30~\cite{Lane:2009cr}, and in other GCs (not analyzed in our work) as well~\cite{Lane:2009cr,Hankey:2010pu,Baumgardt:2009hv,Ibata:2012eq}.

\subsection{Intermediate Mass Black Holes}
Table~\ref{tab:BHbounds} reports the the 95\% C.L. upper limit on $M_{\rm{BH}}$ obtained from the profile likelihood analysis. The limits we find are of the order of (a few times) $10^4\,M_\odot$.

\bigskip

Several methods are available to study IMBHs in GCs, namely to detect the X-ray and radio emission from an IMBHs in the accretion phase, to measure the acceleration of millisecond pulsars in the cluster, or to analyze the dynamics of stars, as performed here. At present no solid evidence has been found from these observations. A tentative detection of an IMBHs of $\sim4\,\times 10^4\,M_{\odot}$ in $\omega$-Centauri has been suggested in~\cite{2008ApJ...676.1008N,2017MNRAS.464.2174B}. This claim has been questioned by subsequent analyses~\cite{vanderMarel:2009mc,2019MNRAS.482.4713Z,2019MNRAS.488.5340B}, with ref.~\cite{vanderMarel:2009mc} obtaining and upper limit on $M_{\rm{BH}}$ of $\sim1.2\,\times 10^4\,M_{\odot.}$
Similarly, the presence for an IMBH of $\sim2\times 10^3\,M_{\odot}$ in 47 Tucanae has been suggested in~\citep{2017Natur.542..203K}, while other works concluded that current data do not favour such interpretation~\cite{Freire:2017mgu,Mann:2018xkm,Abbate:2019qoc,2018MNRAS.481..627A}. An upper limits of $\sim4\,\times 10^3\,M_{\odot}$ has been obtained in~\cite{2018MNRAS.481..627A}. A discussion on the searches for IMBHs in GCs can be found in the recent review~\cite{Greene:2019vlv}.
We notice that our bounds on $M_{\rm{BH}}$ in Table~\ref{tab:chi2} are weaker than the most stringent limits in the literature, which however are derived employing different techniques and observations, more tailored to probe the innermost region of GCs, rather than testing a full mass model as in this work. 

\begin{table}[tb]
\begin{center}
\begin{tabular}{|| c | c ||} 
 \hline
 GC & $M_{BH}^{\rm max}$ [$10^4\,M_{\odot}$]\\
 \hline\hline
 47 Tuc   &2.9 \\ 
 \hline
 NGC 1851  & 2.6\\
 \hline
 NGC 2808  & 4.0 \\
 \hline
 $\omega$-cen & 6.0  \\
 \hline
 M 80   & 3.3 \\
 \hline
 M 22  & 2.2  \\
 \hline
 NGC 6752& 1.2  \\
 \hline
 M 2 &  7.5 \\
 \hline
 M 30 & 2.5  \\
\hline
 NGC 3201 & 1.4  \\
 \hline
\end{tabular}
\end{center}
\caption{95\% C.L. upper limit on the mass of the IMBH.}
\label{tab:BHbounds}
\end{table}
\section{Conclusions}
\label{sec:Conc}
We performed a spherical Jeans analysis of ten GCs of the Milky Way with the goal of providing bounds on the mass components of these objects.
The main contribution to the mass budget is provided by the luminous stellar distribution traced by the surface density profiles.
Additional baryonic dark components can consist of brown dwarfs, neutron stars and stellar-mass black holes. We effectively include these contributions by making the simplifying assumption that they follow the same spatial distribution of the luminous stellar part, and by treating the total stellar mass as a free parameter, i.e., allowing variations of the mass-to-light ratio.  
The main focus of the work was in fact to test the possible presence of additional dark mass terms, in the form of a dark halo, described by an NFW or Burkert mass profile, or an IMBH, described by a point-like mass.

We described the mass distribution in each GC employing a model with 6 parameters pertinent to the stellar profile, 1 parameter providing the stellar mass-to-light ratio, 4 parameters for the velocity anisotropy, and 2 (1) parameters describing DM (IMBH). A different model for the stellar distribution is adopted in Appendix~\ref{sec:King}.
We do not find any statistical preference for DM or IMBH, and derived upper limits on their contribution to the GC mass profile.

Results are shown in Figs.~\ref{fig:upperlimitsDM2}, \ref{fig:upperlimitsDM}, in Table~\ref{tab:chi2} (DM), and in Table~\ref{tab:BHbounds} (IMBH).
The DM mass is typically bounded to be $\lesssim 10^5\,M_\odot$ inside the GC half-light radius. The upper limits on the IMBH mass are at the level of (a few times) $10^4\,M_\odot$.
The results presented in this work can be employed to estimate the $J$- and $D$-factors, i.e., to  set the expected emissions associated to indirect searches of DM.

\acknowledgments
We warmly acknowledge Holger Baumgardt, Sebastian Kamann and Thomas de Boer for providing the datasets used in this work and for useful suggestions. We also thank Mariana Julio, Jarle Brinchmann, Mauro Valli, Eduardo Vitral and Bas Zoutendijk for useful discussions.
We acknowledge support by the PRIN research grant ``From  Darklight  to  Dark  Matter: understanding the galaxy/matter connection to measure the Universe'' No. 20179P3PKJ funded by MIUR, by the research grant ``The Dark Universe: A Synergic Multimessenger Approach No. 2017X7X85'' funded by MIUR, by ``Deciphering the high-energy sky via cross correlation'' and ``Unveiling Dark Matter and missing baryons in the high-energy sky'' funded by the agreement ASI-INAF n. 2017-14-H.0, by the INFN grant ``LINDARK,'' by the ``Department of Excellence" grant awarded by the Italian Ministry of Education, University and Research (MIUR), by the research grant TAsP (Theoretical Astroparticle Physics) funded by Istituto Nazionale di Fisica Nucleare (INFN).

\appendix
\section{LOS velocity dispersion profiles}
\label{sec:AppendixA}

\begin{figure}
\centering
    \begin{subfigure}{1.\linewidth}
    \centering
    \includegraphics[width=1.\textwidth]{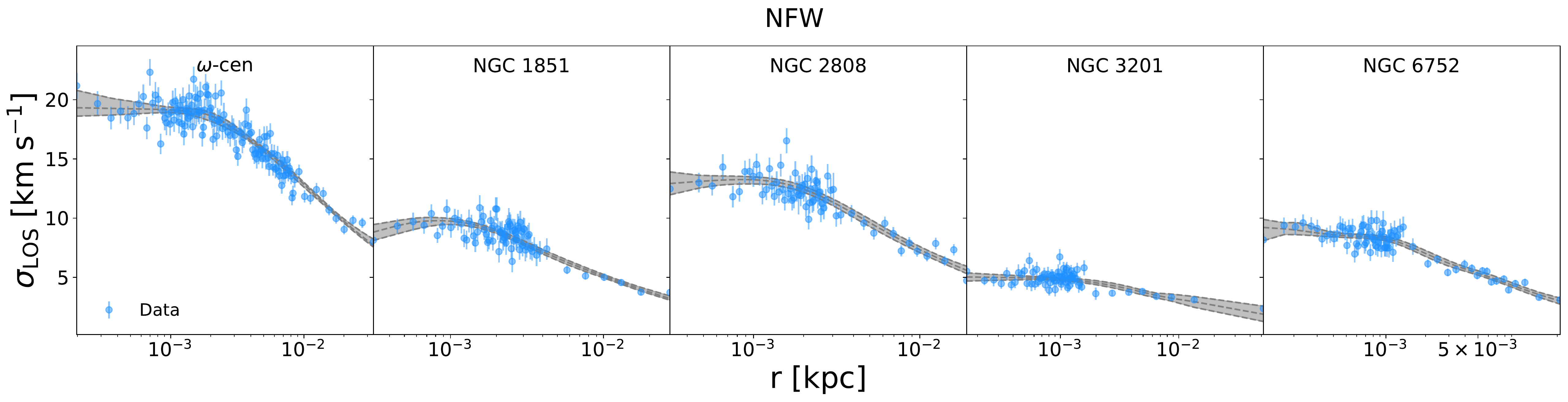}
    \end{subfigure} %

    \hfill

    \begin{subfigure}{1.\linewidth}
    \includegraphics[width=1.\textwidth]{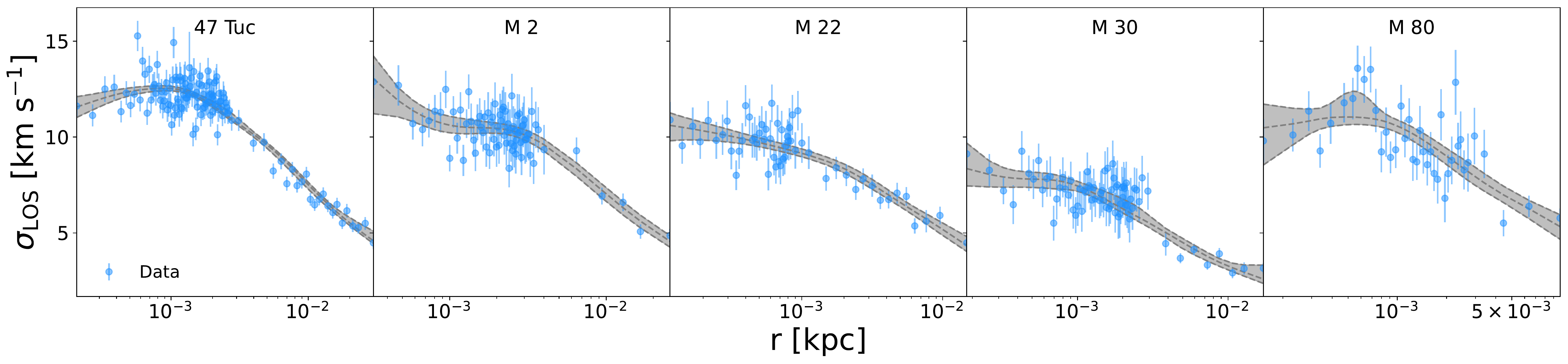}
   \end{subfigure}
   
   \caption{Line-of-sight velocity dispersion for all the GCs considered in this work. Blue points correspond to the kinematic data obtained from the dataset in~\citep{2018MNRAS.473.5591K} and~\citep{2018MNRAS.478.1520B} as explained in Section~\ref{sec:data}. 
   The gray shaded regions show the 95\% intervals of the posterior distributions for the analyses including an NFW DM profile in the mass model.
   }
    \label{fig:sig_los_complete_nfw}

\end{figure}

\begin{figure}
\centering
    \begin{subfigure}{1.\linewidth}
    \centering
    \includegraphics[width=1.\textwidth]{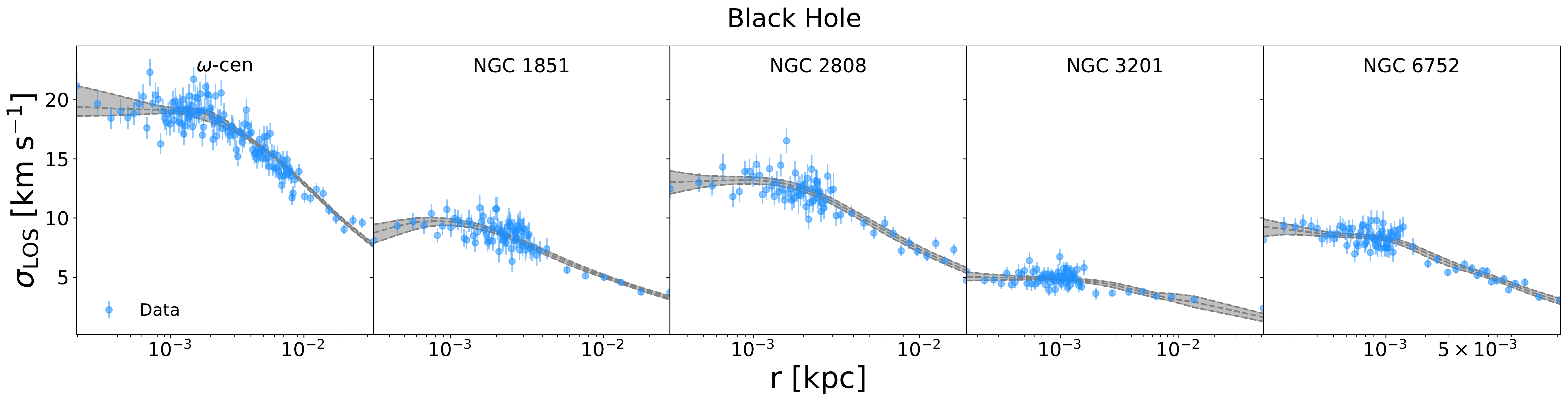}
    \end{subfigure} %

    \hfill

    \begin{subfigure}{1.\linewidth}
    \includegraphics[width=1.\textwidth]{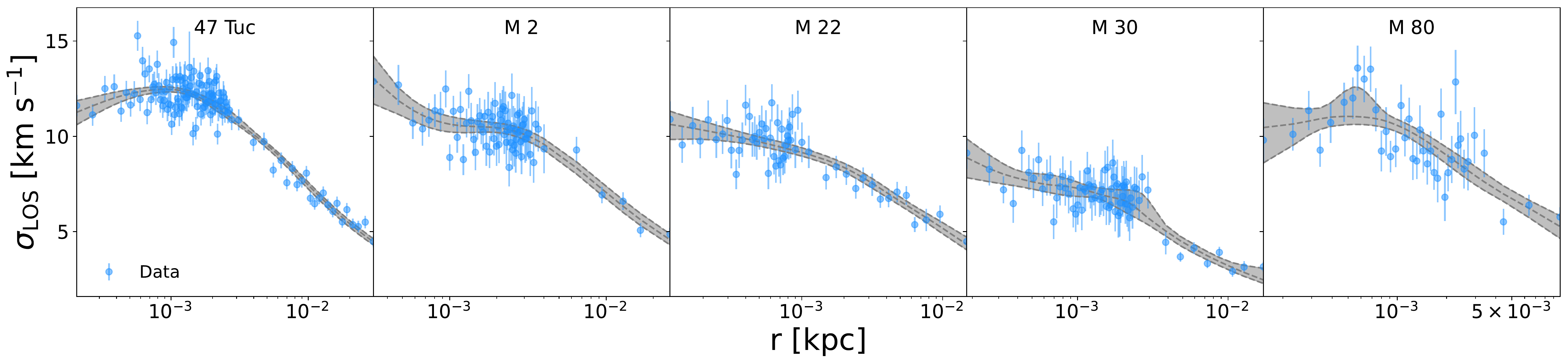}
   \end{subfigure}
   
   \caption{Same as Fig.~\ref{fig:sig_los_complete_nfw} but including an IMBH instead of an NFW DM profile in the analysis.}
    \label{fig:sig_los_complete_bh}

\end{figure}

In this Appendix we present the LOS velocity dispersion profiles of all the analyzed GCs, and the corresponding fits of the model including an NFW DM profile (Fig.~\ref{fig:sig_los_complete_nfw}) or an IMBH (Fig.~\ref{fig:sig_los_complete_bh}).
Points are obtained from the datasets in~\citep{2018MNRAS.473.5591K} and~\citep{2018MNRAS.478.1520B} as explained in Section~\ref{sec:data}.   
The gray shaded regions represent the 95\%C.L. intervals obtained from the posterior distributions.

\begin{figure}
\centering
    \begin{subfigure}{1.\linewidth}
    \centering
    \includegraphics[width=1.\textwidth]{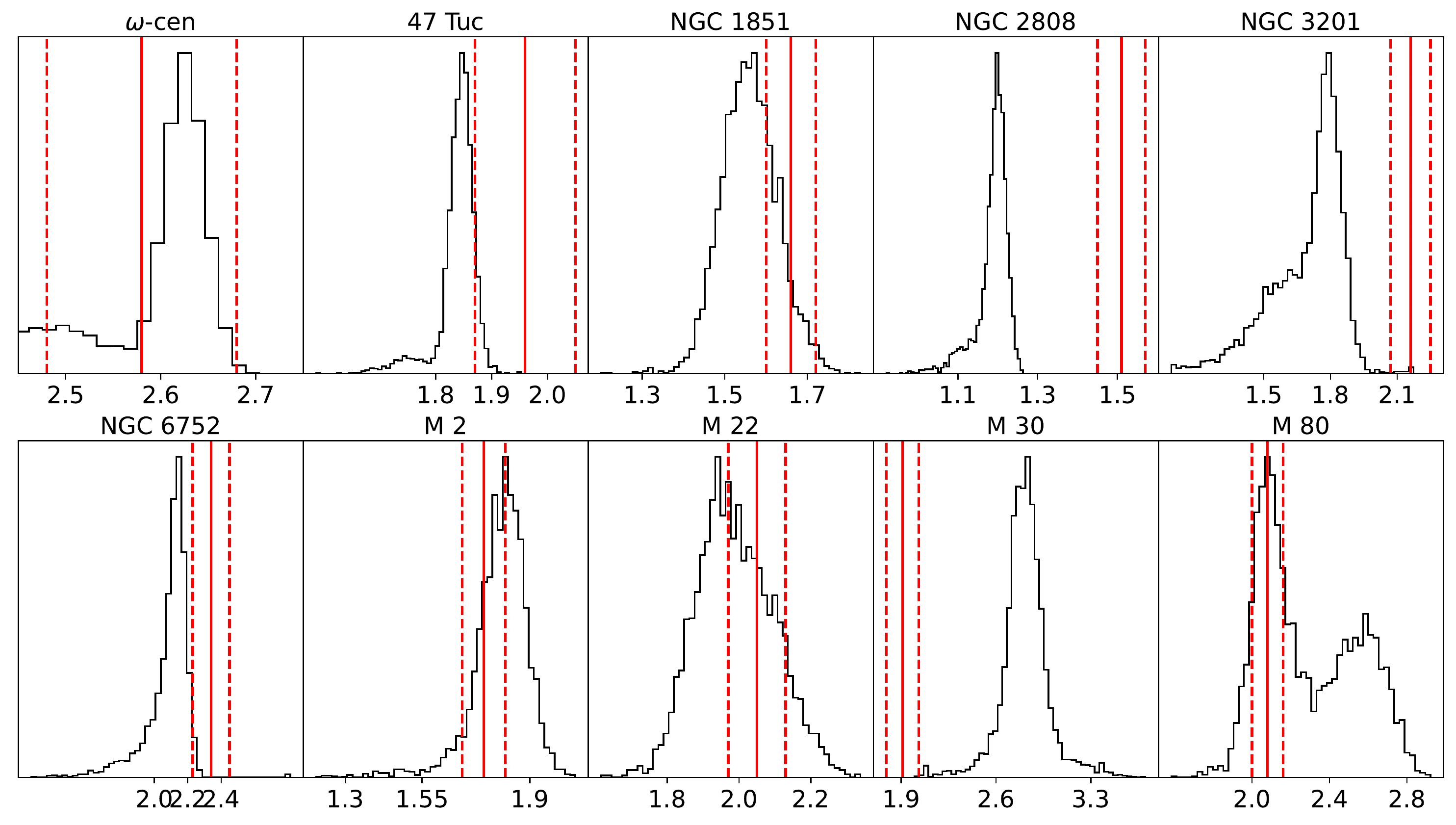}
    \end{subfigure} %

   \caption{Posterior distributions of the Mass-to-Light ratios [$M_\odot/L_\odot$] derived from our analysis compared to the values reported in \citep{hilker_baumgardt_sollima_bellini_2019} (vertical solid  red lines) with their associated 1-$\sigma$ uncertainties (vertical red dashed lines).}
    \label{fig:mass_to_light}

\end{figure}

In Fig.~\ref{fig:mass_to_light} and for all the analyzed GCs, we show instead the posterior distributions of the Mass-to-Light ratios derived from our analysis, compared to the values reported in~\citep{hilker_baumgardt_sollima_bellini_2019}.

\section{Posterior distributions for one example: $\omega$-Centauri}
\label{sec:AppendixB}

\begin{figure}
\centering
    \begin{subfigure}{1.\linewidth}
    \centering
    \includegraphics[width=1.\textwidth]{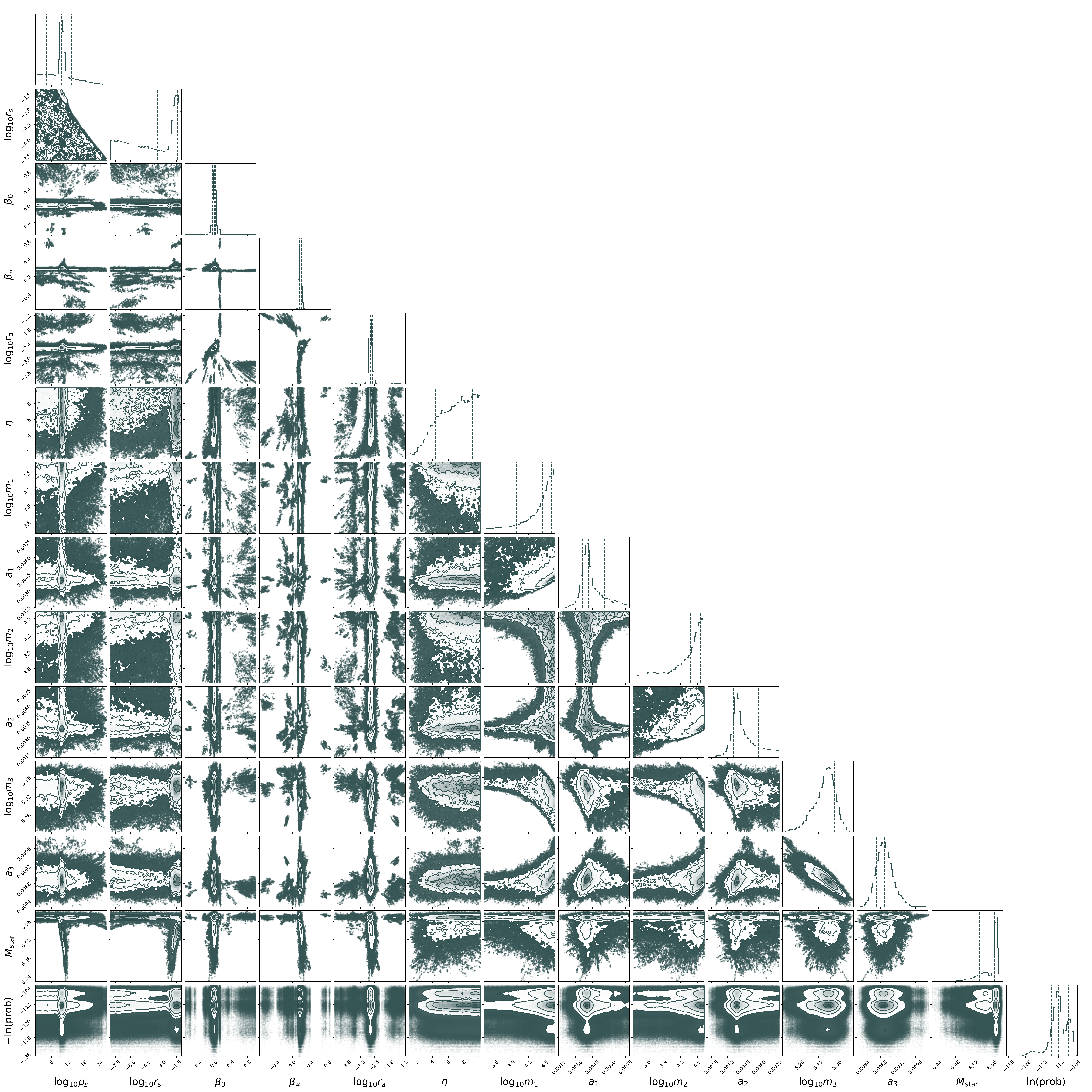}
    \end{subfigure} %

   \caption{Posterior distributions of the different parameters of the model. I. $\log_{10}(\rho_s / [M_\odot \rm{kpc}^{-3}])$. II. $\log_{10}(r_s/\rm{kpc}^{-3})$. III. $\beta_0$. IV. $\beta_{\infty}$. V. $\log_{10}(r_a/\rm{kpc})$. VI. $\eta$. VII. $\log_{10}(m1/M_\odot)$. VIII. $a_1/\rm{kpc}$. IX. $\log_{10}(m2/M_\odot)$. X. $a_2/\rm{kpc}$. XI. $\log_{10}(m3/M_\odot)$. XII. $a_3/\rm{kpc}$. XIII. $\log_{10}(M_{\rm{star}}/M_\odot)$ XIV. $-2\ln(\rm{prob})$   The analysis refers to the $\omega$-Centauri GC and includes an NFW DM profile. }
    \label{fig:triangle_nfw}

\end{figure}

\begin{figure}
\centering
    \begin{subfigure}{1.\linewidth}
    \centering
    \includegraphics[width=1.\textwidth]{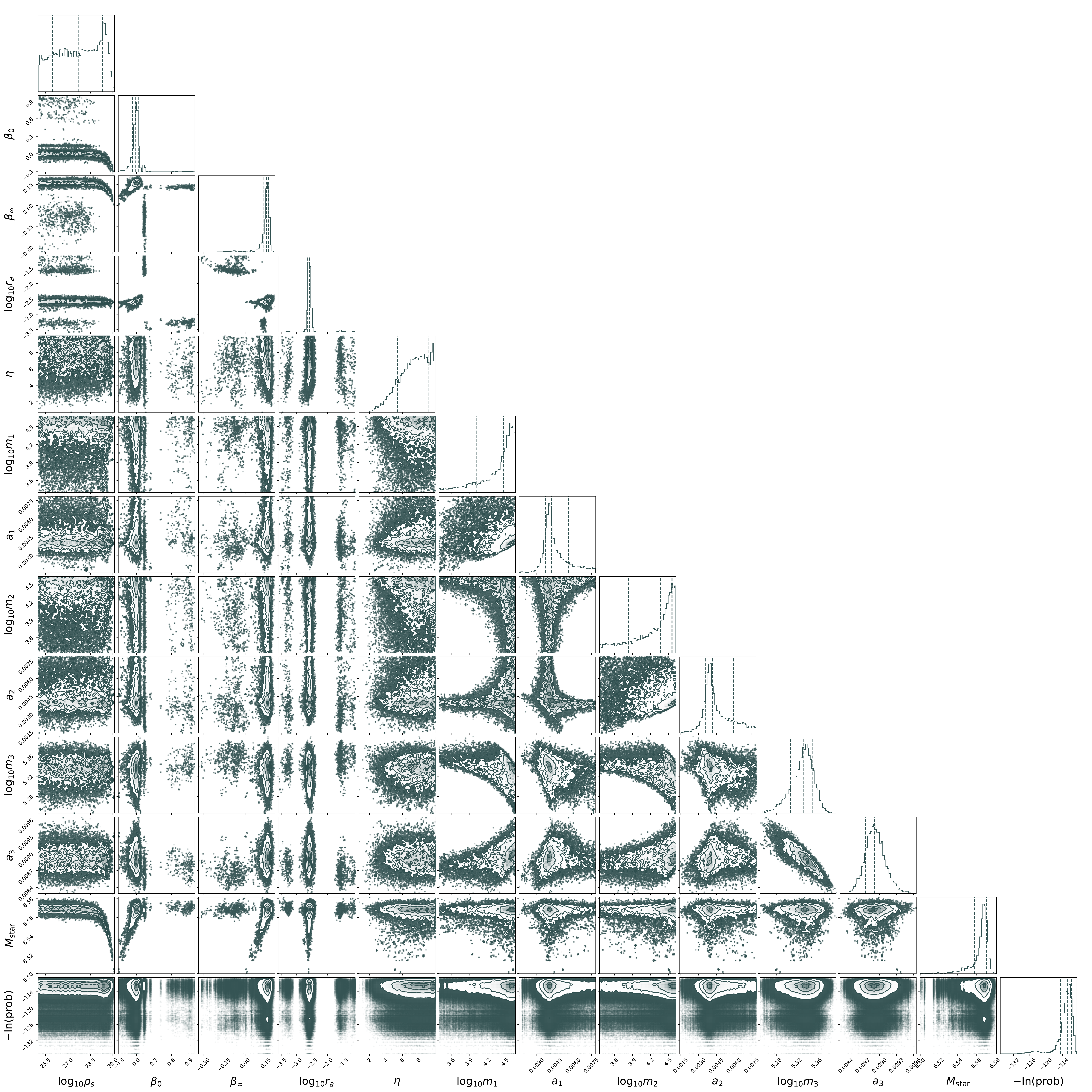}
    \end{subfigure} %

   \caption{Same as in Fig.~\ref{fig:triangle_nfw} but including an IMBH instead of an NFW DM profile in the analysis. Note that we report the value of $\log_{10}(\rho_s/[M_\odot \rm{kpc}^{-3}])$ instead of the IMBH mass. In the sampling we computed the IMBH mass using an NFW profile (Eq. \ref{eq:nfw}) with $r_s=10^{-9}$ kpc and truncated at $R_{\rm half}/100$ (with $R_{\rm half}$ taken from Table~\ref{tab:GC_info}), i.e., $\rho_{{\rm NFW}}(r>R_{\rm half}/100)=0$.
   Then $M_{BH}=4\pi\int_0^{R_{\rm half}/100}\, dr\,r^2\,\rho_{{\rm NFW}}(r,r_s=10^{-9}\,{\rm kpc})$.}
   
    \label{fig:triangle_bh}

\end{figure}

The 1D and 2D posterior distributions of all the parameters considered in the analyses are shown in Fig.~\ref{fig:triangle_nfw} and Fig.~\ref{fig:triangle_bh}, for the cases where the dark component is in the form of an NFW DM profile and an IMBH, respectively. 
These plots refer to one example, the analyses of the $\omega$-Centauri GC.
Other targets can be made available upon request to the authors.

\section{Analysis with the King stellar profile}
\label{sec:King}

\begin{figure}[tb]
\begin{center}
\includegraphics[width=1.\textwidth]{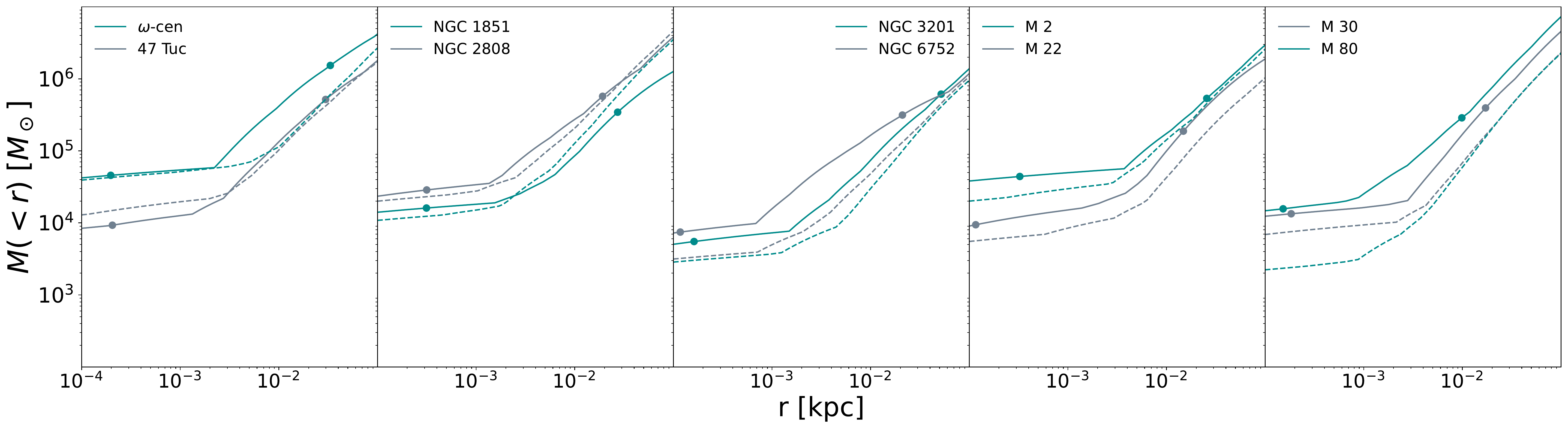}
\caption{Upper limits at 95\% C.L. on the DM mass enclosed within a radius $r$ for the different GCs analyzed and using an NFW DM profile. Solid lines show the bounds obtained using a three Plummer sphere density profile to model the stellar distribution, while the dashed lines are the limits obtained by using a King profile.
Dots correspond to the minimum and maximum radii covered by the measurements of the velocity dispersion.}
\label{fig:upperlimitsDMK} 
\end{center}
\end{figure}

In this section we repeat the analysis described in Section~\ref{sec:data} adopting a different model for the the stellar distribution.
The goal is to study the impact of this ingredient on our results.
We substitute the three Plummer spheres adopted in Section~\ref{sec:data} with a King profile~\cite{1962AJ.....67..471K}. It constitutes a simple, with only two parameters, and physically motivated distribution, often used in the literature. The drawback of its simplicity is that it leads to poor fits to the surface density profiles of the GCs that we have analyzed. In fact, for all the targets, we find a best-fit $\chi^2_{\Sigma_*}$ in \Eq{eq:chi2Sigma} much larger than the one obtained employing three Plummer spheres, and with a reduced $\chi^2_{\Sigma_*}$ value much larger than one. Similar conclusions have been obtained in~\cite{2019MNRAS.485.4906D}.
For this reason, namely, in order to avoid a situation where with a global fit completely dominated by the surface density part and with kinematic data having little impact, we slightly modify the procedure discussed in Section~\ref{sec:data}. For each GC, the parameters of the King model are fixed to the values which minimize $\chi^2_{\Sigma_*}$ and the Mass-to-Light ratio is fixed to a physically motivated value (taken from the best-fit of the analysis in the main text). Then, we exclude the surface density data from the likelihood in~\Eq{eq:like}, including only the measurements of the LOS velocity dispersion and the virial shape parameters ($\chi^2=\chi^2_{\rm LOS}+\chi^2_{\rm VSP1}+\chi^2_{\rm VSP2})$.
This procedure is less general than the one of the main text, but offers an alternative way to constrain the dark component using a physically motivated model.

The results of the analysis including a NFW DM halo are presented in Fig.~\ref{fig:upperlimitsDMK}, where we show the 95\%C.L. upper limits on the DM mass enclosed within a given GC radius $r$. We do not find statistical preference for a DM  component in any of the GCs analyzed. With respect to the bounds obtained in Section~\ref{sec:darkmatter}, by modeling the stellar profile with three Plummer spheres, we see that, in the case of the King profile, limits are typically similar or slightly stronger. The latter can be ascribed to the fact that here we are adopting a more constrained statistical procedure.

\clearpage

\bibliographystyle{hunsrt}
\bibliography{bibliography}

\end{document}